\documentclass[pre,twocolumn,showpacs,floatfix]{revtex4}
\usepackage{graphicx}
\usepackage{amsmath}
\newcommand \etal {{\it et al.\ }}
\newcommand \be {\begin{equation}}
\newcommand \ee {\end{equation}}
\newcommand \ben {\begin{eqnarray}}
\newcommand \een {\end{eqnarray}}

\begin{document}

\title{Eighth-order phase-field-crystal model for two-dimensional crystallization}
\author{A. Jaatinen$^{1,2}$ and
T. Ala-Nissila$^{1,3}$}

\affiliation{$^1$Department of Applied Physics, Aalto University
School of Science, P.O. Box 11000, FI-00076 Aalto, Finland}
\affiliation{$^2$Department of Materials Science and Engineering,
Aalto University School of Science, P.O. Box 16200, FI-00076 Aalto,
Finland} \affiliation{$^3$Department of Physics, Brown University,
Providence RI 02912-1843}

\date{19 November 2010}

\begin{abstract}
We present a derivation of the recently proposed eighth order phase
field crystal model [Jaatinen \etal, Phys. Rev. E {\bf 80}, 031602
(2009)] for the crystallization of a solid from an undercooled melt.
The model is used to study the planar growth of a two dimensional
hexagonal crystal, and the results are compared against similar
results from dynamical density functional theory of Marconi and
Tarazona, as well as other phase field crystal models. We find that
among the phase field crystal models studied, the eighth order
fitting scheme gives results in good agreement with the density
functional theory for both static and dynamic properties, suggesting
it is an accurate and computationally efficient approximation to the
density functional theory.
\end{abstract}


\maketitle

\section{Introduction}

Understanding crystal formation from an undercooled melt is of
significant academic interest due to the complex phenomena involved
in crystallization, and also of practical interest due to its
relevance to a vast amount of industrial processes. During the past
decade, rapidly evolving progress in microscopic understanding of
phenomena involved in solidification has followed the introduction
of the phase field crystal (PFC) model \citep{elder2002,elder2004}.
This model was first introduced as a phenomenological extension of
the traditional phase field models \citep{emmerich2008} such that
the order parameter field exhibits the crystalline nature of the
underlying crystal lattice. The most significant advantage of PFC
type of models over the traditional phase field models is that
including the periodic structure of the field in the model will
naturally result in inclusion of many crystal structure related
properties, such as elasticity, plasticity and grain boundaries
\citep{elder2002}. Since its introduction, the PFC model has been
applied to modeling elastic and plastic deformation of materials
\citep{elder2004,stefanovic2006}, dislocation dynamics
\citep{berry2006}, crystal growth \citep{tegze2009,teeffelen2009},
static and dynamic properties of driven two dimensional overlayers
\citep{achim2006,achim2008,achim2009}, etc.

Because of the periodic nature of the order parameter field in the
PFC model, it is not hard to come up with an intuitive
interpretation that the field must be related to the atomic number
density of the underlying system. On the other hand, studies of
classical density functional theory (DFT), most commonly in the
context of inhomogenous liquids \citep{hansen2006}, have aimed at a
microscopic derivation of the static (and more recently dynamic
\citep{marconi1999}) properties of the systems under study by using
the microscopic density as a field variable in the theory. The
extension of this approach to crystallization is known as the DFT of
freezing, which has, in its many forms, been applied to study
freezing of many different classical systems with a varying degree
of success \citep{ramakrishnan1979,singh1991}. In 2007, Elder \etal
\citep{elder2007} introduced the idea that assuming the field under
study in the PFC model to be linearly proportional to the atomic
number density in the DFT, the PFC model can be viewed as a
simplified version of the DFT, and showed that the free energy
functional used in PFC studies can be derived from the DFT by making
certain approximations. Wu and Karma \citep{wu2007} introduced
another way of obtaining the parameters for the PFC model using a
DFT-like approach. In a recent paper, the strengths and weaknesses
of the approaches proposed in Refs. \citep{elder2007} and
\citep{wu2007} were studied, and a new variant of the PFC model
known as the eighth-order fitting model (EOF), which reproduces
certain thermodynamic properties of the material under study
significantly more accurately than the previously proposed methods,
was proposed \citep{jaatinen2009}. More recently, the EOF model has
been applied to study grain boundaries \citep{jaatinen2010} and
homogenous nucleation \citep{toth2010} of body-centered cubic iron.

In the present work, we will present an alternative interpretation
of the EOF model, in which the field under study in the EOF is
related to the physical atomic number density through a convolution,
which filters out the sub-atomic wavelength Fourier modes from the
atomic number density. Using this interpretation, we are able to
derive the free energy of the EOF model in a way which we believe is
more consistent with the original DFT than the previously presented
derivations. Predictions of the EOF model are then tested against
the DFT and other related PFC models. While most studies of the PFC
model's connection to DFT have concentrated on the free energy, i.e.
the static properties of the model, we will also compare the
predictions for crystal growth rates, which is a dynamical
phenomenon. Similar comparison between crystal growth rates in DFT
and PFC models has previously been published by van Teeffelen \etal
who found the growth rates of colloidal crystals in the early stages
of solidification agree relatively well between the DFT and their
choice of PFC models (not including the EOF) \citep{teeffelen2009}.
In the present work, we aim at a more thorough assessment of the
crystal front propagation in the DDFT and PFC models. Instead of the
initial growth rate, we aim at assessing the steady state front
propagation velocity in both the diffusion controlled and interface
kinetics controlled regimes using both DDFT and EOF models. In
addition, the results from these models are compared to results of a
more traditional fourth-order PFC approach, and the "PFC1" model
that van Teeffelen \etal proposed, and argued to be a more accurate
approximation to DFT than the other model utilized in their
comparison \citep{teeffelen2009}. We find that among the PFC models
studied, the EOF gives results in best agreement with the DFT for
both static properties and crystal growth.

\section{Theory}
\label{sec:theory}

We study solidification dynamics in a two dimensional ensemble of
Brownian particles interacting via an inverse twelfth-power pair
potential,
\begin{equation}
\label{eq:potential} v(r)=\epsilon \left( \frac{\sigma}{r}
\right)^{12},
\end{equation}
where $r$ is the interparticle separation, $\epsilon$ sets the
energy scale and $\sigma$ is the diameter of the particles. Due to
scaling properties, all structural and thermodynamic properties of
this model system only depend on the scaled density
\begin{equation}
\label{eq:density} \tilde{\rho} = \left( \frac{\epsilon}{k_B T}
\right)^{\frac{1}{6}} \rho \sigma^2,
\end{equation}
where $k_BT$ is the thermal energy scale and $\rho=N/A$ is the
number of particles per unit area. According to molecular dynamics
simulations of Broughton \emph{et al.}, the equilibrium state of the
system is a fluid at densities up to $\tilde{\rho}_l=0.987$, and a
hexagonal solid at densities above $\tilde{\rho}_s=1.006$, while
between these two densities the equilibrium state is a coexistence
of the solid and liquid phases \citep{lowen1996,broughton1982}. As
we assume that hydrodynamic interactions and the inertial terms can
be neglected, the equations of motion for the particles are given by
\begin{equation}
\label{eq:brown} \dot{\bf r}_i = \gamma^{-1} ({\bf F}_i + {\bf
f}_i),\ i=1 \ldots N,
\end{equation}
where the dot denotes time derivative, $\gamma$ is a friction
coefficient, ${\bf F}_i$ is the force from the other particles and
an external field acting on particle $i$, and ${\bf f}_i$ is a
Gaussian random force that fulfills the fluctuation-dissipation
theorem.

\subsection{Dynamical density functional theory}

In the DDFT approach, instead of solving the positions of individual
particles as a function of $t$, one derives an equation of motion
for the one-particle density defined by
\begin{equation}
\label{eq:rho} \rho ({\bf r},t) = \sum_i \left< \delta ({\bf r}-{\bf
r}_i(t) ) \right>,
\end{equation}
where the angular brackets denote a noise average
\citep{teeffelen2009}. Marconi and Tarazona \citep{marconi1999} have
shown that from the equations of motion \eqref{eq:brown}, one can
derive an equation of motion for $\rho({\bf r},t)$ through a
coordinate transformation and a subsequent noise averaging. In
another procedure, Archer and Evans \citep{archer2004} have derived
the same equation of motion by using the Smoluchowski equation as
their starting point. The equation of motion for $\rho({\bf r},t)$
resulting from both of these derivations reads
\begin{equation}
\label{eq:ddfteqmotion} \dot{\rho}({\bf r},t)= \gamma^{-1} \nabla
\cdot \left( \rho({\bf r},t) \nabla \left( \frac{\delta  F \left[
\rho({\bf r},t) \right] }{\delta \rho({\bf r},t) } \right) \right),
\end{equation}
where $ F \left[ \rho({\bf r},t) \right]$ is the Helmholtz free
energy of the system described by a density field $\rho({\bf r},t)$.
As noted in the recent work of Ramos \etal  this equation of motion
can also be obtained in the overdamped limit of a more general
equation of motion for the number and momentum densities, if the
effective Hamiltonian is replaced by the free energy and thermal
fluctuations are ignored \citep{ramos2010}.

The free energy $F$ consists of three parts, $ F = F_{id}+ F_{ex}+ F
_{xs}$, where the first term represents the ideal gas contribution,
\begin{equation}
\label{eq:Fid}  F_{id}=k_BT \int d{\bf r} \rho ({\bf r}) \left(
\ln(\rho ({\bf r}) \lambda_T^2)-1 \right),
\end{equation}
where $\lambda_T$ is thermal de Broglie wavelength and the second
term is a contribution from an external field,
\begin{equation}
\label{eq:Fex}  F_{ex}= \int d{\bf r} \rho ({\bf r}) u_{ex}({\bf
r}),
\end{equation}
where $u_{ex}({\bf r})$ is an external field acting on the
particles. The third part of $F$ is the excess, which is due to the
interparticle interactions. For this quantity, exact expressions
only exist for a very limited range of cases, and more generally,
approximations will have to be made \citep{hansen2006}. In the
present work, we will use the simplest possible non-local
approximation that is an expansion of $F_{xs}$ in powers of $\Delta
\rho = \rho-\rho_0$ around a uniform reference density $\rho_0$,
where
\begin{equation}
\begin{split}
\label{eq:Fxs} F_{xs}  = & -k_BT \int d{\bf r} c^{(1)}(\rho_0)\Delta
\rho ({\bf r})
\\  - & \frac{k_BT}{2} \iint d{\bf r} d{\bf r}' \ \Delta \rho ({\bf
r}) c^{(2)}(|{\bf r}-{\bf r}'|,\rho_0) \Delta \rho ({\bf r}'),
\end{split}
\end{equation}
where $c^{(n)}$ are called $n^{\mathrm{ th}}$ order direct
correlation functions. The quantity $c^{(2)}$ is the
Ornstein-Zernike direct correlation function that can be obtained
from experiments, computer simulations or a number of approximate
closure relations to the Ornstein-Zernike equation
\citep{hansen2006}. In the present work, we will utilize the
well-known Percus-Yevick closure relation with the pair potential
Eq. \eqref{eq:potential} to obtain $c^{(2)}$. The reference density
$\rho_0$ is chosen such that $F$ has two equal minima: The trivial
uniform minimum $\rho({\bf r})=\rho_0$ and another where $\rho({\bf
r})$ has a hexagonal structure (the external field $u_{ex}$ is set
to zero). This is the procedure taken in most DFT studies of
freezing, and the resulting $\rho_0$ is interpreted as the freezing
point of the liquid. Indeed, the free energy functional defined by
Eqs. \eqref{eq:Fid} and \eqref{eq:Fxs} is the simplest free energy
functional used in static DFT studies of freezing, and its success
has varied from case to case (\citep{singh1991} and references
therein). The free energy defined by Eqs. \eqref{eq:Fid} and
\eqref{eq:Fxs} is also the free energy functional to which Elder
\emph{et al.} attempted to connect the free energy used in PFC
studies \citep{elder2007}.

Putting our free energy functional (with $u_{ex}=0$) together with
Eq. \eqref{eq:ddfteqmotion}, and rescaling the density field
variable as $\rho({\bf r},t) = \rho_0 (1+n({\bf r},t))$, we end up
with an equation of motion
\begin{equation}
\label{eq:ddftn} \frac{ \partial n}{\partial \tau} = \nabla^2 n -
\nabla \cdot \left( (1+n) \nabla \int d{\bf r}' C(|{\bf r}-{\bf
r}'|) n({\bf r}') \right),
\end{equation}
where the rescaled time $\tau =\gamma^{-1} k_BT t$ and
$C=\rho_0c^{(2)}$. Very few studies of the dynamics of
crystallization using Eq. \eqref{eq:ddftn} are found in the
literature. Van Teeffelen \etal have studied the dynamics of
colloidal crystal nucleation and found results that seemed to agree
well with the results of molecular dynamics simulations
\citep{teeffelen2007}. The same group has studied the initial growth
velocity of a colloidal crystallization front using Eq.
\eqref{eq:ddftn} and compared the results against the results
obtained from the PFC model \citep{teeffelen2009}. To our knowledge,
no other attempts to assess solidification dynamics by direct
application of Eq. \eqref{eq:ddftn} exist in the literature.

\subsection{Eighth order phase field crystal model}

The eighth order phase field crystal model (EOF) was recently
presented in Ref. \citep{jaatinen2009} in the context of
quantitative modelling of body-centered cubic iron. In that case,
the model was shown to reproduce the anisotropic solid-liquid
interfacial free energies, bulk moduli of solid and liquid phases,
and the equilibrium coexistence gap between them to a quantitatively
satisfactory precision. While some of these properties had been
reproduced in previous versions of phase field crystal models, the
combination seemed inaccessible without the eighth order extension.
\citep{jaatinen2009}. In a subsequent study, it was shown that the
EOF is also capable of describing grain boundary energies of bcc
iron quantitatively \citep{jaatinen2010}.

Despite its quantitative success, there remain open issues in the
EOF model. First, the previously presented derivation of the EOF is
based on the assumption that the field $n$, related to the atomic
number density as in the case of DFT, is small, such that the
logarithmic term in the free energy could be expanded as a power
series. However, it is well known that the actual atomic number
density in the solid resembles a set of highly localized Gaussians,
which does not agree with the assumption of $n$ being small. Neither
does it suggest that the non-local part of the free energy could be
assumed local in $k$-space, as must be assumed, when expanding the
direct correlation function in $k$-space around its main maximum.
The equilibrium $n$-field resulting from these approximations is
highly localized in $k$-space, having little resemblence to the
Gaussian-like density obtained from the DFT. Most prominently, in
the supposedly empty spaces between the lattice sites, the field $n$
resulting from the EOF reaches values smaller than $-1$,
corresponding to unphysical negative densities.

In what follows, we will present an alternative derivation of the
EOF model which, although far from exact, avoids the previously
mentioned caveats. The key to our derivation is the obvious fix in
the interpretation of the field $n$ used as the field variable in
the EOF model: Instead of insisting that the field $n$ in the EOF
model would be locally and linearly related to $\rho$ in the same
way as in the DDFT model, we will assume they are related through a
weighing function $w$ as
\begin{equation}
\label{eq:EOFn}
n({\bf r})= \rho_0^{-1} \int d{\bf r}' w(|{\bf r}-{\bf r}'|) (\rho({\bf r})-\rho_0).
\end{equation}
In the current approach, the weighing procedure introduced in Eq.
\eqref{eq:EOFn} will act as a Fourier filter, cutting off the short
wavelength modes of $\rho$ that have very small amplitude in the
periodic solutions of the PFC model. In other words, the field $n$
in the bulk of the solid phase will closely resemble the one-mode
approximation,
\begin{equation}
\label{eq:nonemode}
\begin{split}
n({\bf r}) & \approx  n_0 + \\ & 2u \left( \cos \left( \frac{4 \pi y}{\sqrt{3}d_{nn}}\right)-2\cos\left( \frac{2 \pi x}{d_{nn}}\right) \cos\left( \frac{2 \pi y}{\sqrt{3}d_{nn}}\right) \right),
\end{split}
\end{equation}
where $n_0$ is the fractional density change and $d_{nn}$ is the
nearest-neighbor distance, even though the underlying density field
were highly peaked around lattice sites.

A convenient choice for $w$ is a function, whose Fourier transform is given by
\begin{equation}
\label{eq:wk}
\hat{w}(k) = \sqrt{\frac{1-\hat{C}(k)}{1-\hat{C}_{EOF}(k)}}
\end{equation}
where $C_{EOF}(k)$ is the "approximation" to $C(k)$ introduced in
\citep{jaatinen2009},
\begin{equation}
\label{eq:cexpansion} \hat{C}_{EOF}(k) = \hat{C}(k_m) - E_S \left(
\frac{k_m^2 - k^2}{k_m^2} \right)^2 - E_B \left( \frac{k_m^2 -
k^2}{k_m^2} \right)^4,
\end{equation}
where $k_m$ is the position of the main peak in $\hat{C}(k)$, $E_S$
is chosen such that second derivatives at the peak of the original
and the approximated curves are equal, and $E_B$ is then chosen such
that the infinite-wavelength ($k=0$) limits are equal. The function
$\hat{C}_{EOF}(k)$ will follow the original $\hat{C}(k)$ very
closely from the $k=0$ limit up to the main peak at $k_m$, after
which the two curves diverge, $\hat{C}(k)$ approaching zero in an
oscillatory fashion, while $\hat{C}_{EOF}(k)$ falls in the negative
infinity, such that $\hat{w}(k)$ defined by Eq. \eqref{eq:wk} will
fall close to zero rapidly after the main peak, providing the
desired Fourier filter property mentioned earlier.

What makes the choice of $\hat{w}(k)$ defined by Eq.
\eqref{eq:cexpansion} particularly convenient is that the linear
part of the free energy (linearization of Eqs. \eqref{eq:Fid} and
\eqref{eq:Fxs}) can now be exactly written as
\begin{equation}
\label{eq:eoflinear}
\begin{split}
\frac{\beta F_{lin}}{\rho_0} = & \frac{\rho_0^{-2}}{2} \iint d{\bf r} d{\bf r}' \ \Delta \rho ({\bf
r}) \left( \delta({\bf r}-{\bf r}') - C({\bf r}-{\bf r}') \right) \Delta \rho ({\bf r}')\\
 = &  \frac{1}{2} \int d{\bf r} n({\bf r}) \left( 1-C_{EOF}(\nabla^2) \right) n({\bf r}),
\end{split}
\end{equation}
where $\delta$ is the Dirac delta function and $C_{EOF}(\nabla^2)$
is the inverse Fourier transform of $\hat{C}_{EOF}(k)$, i.e. Eq.
\eqref{eq:cexpansion} with $-k^2$ replaced by $\nabla^2$.

Unfortunately for the non-linear parts of the free energy,
\begin{equation}
\begin{split}
\label{eq:FnlDFT}
\beta F_{nl} = & \beta(F-F_{lin}) \\
 = & \int d{\bf r}\left( \rho ({\bf r})
\ln(\rho ({\bf r}) / \rho_0 )- \Delta \rho({\bf r}) - \frac{\Delta \rho({\bf r})^2}{2 \rho_0} \right),
\end{split}
\end{equation}
the situation is not nearly as trivial, because the expressions will
become both non-local and non-linear. However, as we know that
$F_{lin}$ defined by Eq. \eqref{eq:eoflinear} already provides us
with both a preferred wavelength of fluctuations in the system, and
a large free energy penalty for Fourier modes with $k>>k_m$, we
argue that if the amplitudes $u$ in Eq. \eqref{eq:nonemode} vary on
length scales larger than the range of the weighing function, it may
be sufficient to aproximate $F_{nl}$ with a functional that is local
in terms of the field $n$. To fit this purpose, we postulate a
simple, local, non-linear functional, consisting of third and fourth
order terms,
\begin{equation}
\label{eq:FnlEOF}
\frac{\beta F_{nl,EOF}}{\rho_0} = \int d{\bf r} \left( -\frac{a}{6} n({\bf r})^3 + \frac{b}{12} n({\bf r})^4 \right),
\end{equation}
where $a$ and $b$ are phenomenological constants. As a consequence
of this crude approximation, ignoring practically all information
about the Fourier modes with $k>>k_m$, it is admittedly obvious that
the density field obtained from a solution of the EOF model through
inverting Eq. \eqref{eq:EOFn} will not be an accurate approximation
to the real underlying density field, unless a more accurate
approximation to $F_{nl}$ will be presented in subsequent studies.

It should be noted that the free energy functional for the EOF
obtained by summing up Eqs. \eqref{eq:eoflinear} and
\eqref{eq:FnlEOF} is exactly the same as used in previous EOF
studies \citep{jaatinen2009,jaatinen2010}. One crucial difference,
in addition to the different approximations involved, is that the
current derivation does not suggest that $a=b=1$, like the previous
version, based on Taylor expansion, did. \citep{jaatinen2009}
Instead, in the current approach, it is clear that the proper way to
choose parameters $a$ and $b$ is such that free energies of relevant
density profiles are reproduced as accurately as possible (in the
end, to correct for the flaws in the derivation, fitting the
parameters $a$ and $b$ with the desired amplitude of the solid phase
was the approach taken in the previous studies as well). In order to
gain insight into how these parameters should be chosen, consider a
density field consisting of an infinite set of normalized Gaussians
in a triangular lattice,
\begin{equation}
\label{eq:gaussians}
\rho ({\bf r}) = \sum_i \frac{\alpha}{\pi} e^{- \alpha |{\bf r}-{\bf R}_i|^2},
\end{equation}
where ${\bf R}_i$'s are positions of lattice points that belong in
the underlying hexagonal lattice. It is well known that in the bulk
of the solid, this is a fairly accurate approximation to the density
profile that results from DFT of freezing \citep{singh1991}. More
specifically, consider the case when the density of lattice points
$2/(\sqrt{3}d_{nn}^2)=\rho_0$ and the length of the principal
reciprocal lattice vector $|{\bf G}_m|=4 \pi / (\sqrt{3}d_{nn})$
coincides with position of the main peak in $\hat{C}$, i.e. $|{\bf
G}_m|=k_m$. Then, through Eqs. \eqref{eq:EOFn} and \eqref{eq:wk},
the field $n$ will be closely approximated by Eq.
\eqref{eq:nonemode}, with $n_0=0$, and the amplitude $u$ is related
to the Gaussian parameter $\alpha$ as
\begin{equation}
\label{eq:gaussu}
u=e^{- \frac{|{\bf G}_m|^2}{2 \alpha}}.
\end{equation}
It is then straightforward to calculate from Eqs.
\eqref{eq:eoflinear} and \eqref{eq:FnlEOF} that free energy of the
system described by this single-mode $n$ field will be given by
\begin{equation}
\label{eq:FuEOF}
\frac{\beta F_{EOF}}{N} = 3 (1-\hat{C}(k_m))u^2 - 2 a u^3 + \frac{15}{2} b u^4.
\end{equation}
In what follows, this expression will be related to the small and
large $\alpha$ limits of the original free energy, given by Eqs.
\eqref{eq:Fid} and \eqref{eq:Fxs}, in the Gaussian approximation.

In the limit of small $\alpha$, deviations from uniformity in the
density field are small, and therefore, instead of the full free
energy functional, it is sufficient to consider the linearised
version defined by Eq. \eqref{eq:eoflinear},
\begin{equation}
\label{eq:smallalpha}
\begin{split}
\frac{\beta F}{\rho_0} \approx & \frac{\rho_0^{-2}}{2} \iint d{\bf r} d{\bf r}' \ \Delta \rho ({\bf
r}) \left( \delta({\bf r}-{\bf r}') - C({\bf r}-{\bf r}') \right) \Delta \rho ({\bf r}')\\
 = &  \frac{1}{2} \sum_{i} \left( 1-\hat{C}(|{\bf G}_i|) \right) e^{- \frac{|{\bf G}_i|^2}{ \alpha}}
\end{split}
\end{equation}
where the sum is over all non-zero reciprocal lattice vectors. Using
Eq. \eqref{eq:gaussu}, this can be re-written in terms of $u$ as
\begin{equation}
\label{eq:smallalphau}
\begin{split}
\frac{\beta F}{\rho_0} \approx
 \frac{1}{2} \sum_{i} \left( 1-\hat{C}(|{\bf G}_i|) \right) u^{2 \left( \frac{|{\bf G}_i|}{|{\bf G}_m|}\right)^2}.
\end{split}
\end{equation}
Now it is easily observed that in the limit where $u$ is small, the
leading contribution to the sum in Eq. \eqref{eq:smallalphau} comes
from the shortest reciprocal lattice vectors, for which $|{\bf
G}_i|=|{\bf G}_m|$. As the number of vectors in this first star of
reciprocal lattice vectors is six, we obtain exactly the same
small-$u$ behaviour from Eqs. \eqref{eq:FuEOF} and
\eqref{eq:smallalphau}. This result may not be surprising, due to
the exact relation between the linear parts of free energies in the
models.

In the limit of large $\alpha$, i.e. highly localised density peaks,
the free energy can be accurately evaluated by ignoring the overlap
between peaks in the ideal gas term \citep{singh1991}, resulting in
\begin{equation}
\label{eq:largealpha} \frac{\beta F}{N} \approx \ln \left(
\frac{\alpha}{\pi \rho_0} \right)-1-\frac{1}{2} \sum_{i}
\hat{C}(|{\bf G}_i|) e^{- \frac{|{\bf G}_i|^2}{ \alpha}}.
\end{equation}
Using Eq. \eqref{eq:gaussu}, this can also be expressed as a function of $u$ as
\begin{equation}
\label{eq:largealphau} \frac{\beta F}{N} \approx  \ln \left( \frac{2
\pi}{\sqrt{3}e} \right)-\ln(-\ln(u)) -\frac{1}{2} \sum_{i}
\hat{C}(|{\bf G}_i|) u^{2 \left( \frac{|{\bf G}_i|}{|{\bf
G}_m|}\right)^2}.
\end{equation}

\begin{figure}
\begin{center}

{\includegraphics[width=86mm]{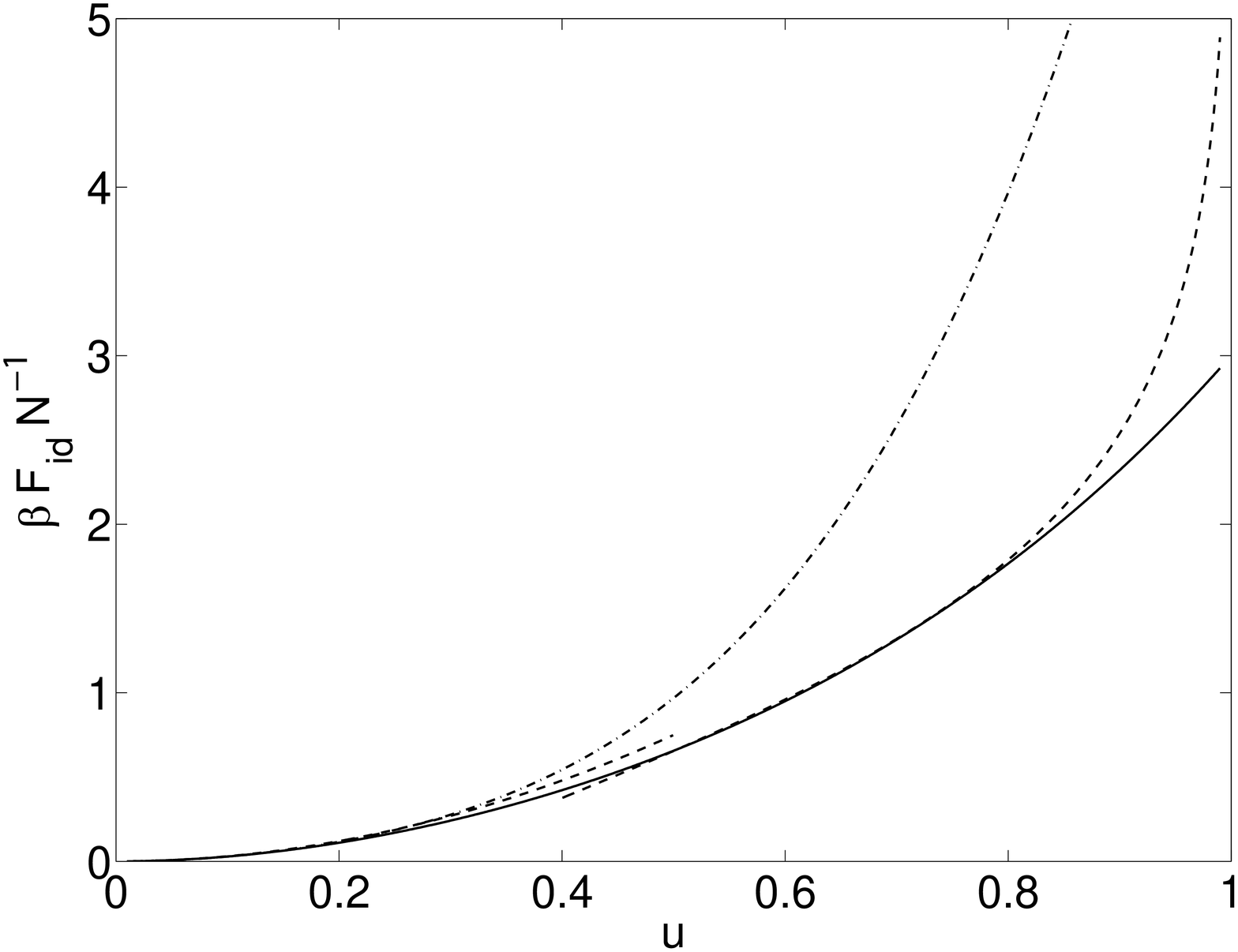}}

\caption{\label{fig:Fu} Ideal free energy as a fucntion of $u$ in
the one-mode approximation to the EOF model, as indicated in Eq.
\eqref{eq:idfit} with $a_{id}=3/4$ and $b_{id}=1/5$ (solid line).
Dashed lines represent the small- and large-$u$ limits obtained from
the original $F_{id}$ in the Gaussian approximation. For comparison,
the dashed-dotted line shows the curve expected from a Taylor
expansion of $F_{id}$ (i.e. $a_{id}=b_{id}=1$) which is seen to
significantly overestimate $F_{id}$.}

\end{center}
\end{figure}

We note that it might be possible to come up with a form of $F_{nl,EOF}$ that would resemble this form more closely than the simple fourth order polynomial in Eq. \eqref{eq:FuEOF}. For the time being, however, we find it sufficient to study the implications of Eq. \eqref{eq:largealphau} to the parameters $a$ and $b$ of Eq. \eqref{eq:FnlEOF}. Perhaps the most interesting of these implications is the contribution of the ideal gas term in the parameters $a$ and $b$. Ignoring all terms that are proportional to $\hat{C}(k)$ in Eqs. \eqref{eq:FuEOF} and \eqref{eq:largealphau} leaves us with the problem of finding coefficients $a_{id}$ and $b_{id}$ (subscript $id$ refers to the ideal free energy), such that
\begin{equation}
\label{eq:idfit}
3 u^2 - 2 a_{id} u^3 + \frac{15}{2} b_{id} u^4 \approx  \ln \left( \frac{2 \pi}{\sqrt{3}e} \right)-\ln(-\ln(u)).
\end{equation}
There are, of course, an infinite number of ways to perform this
fit. After experimenting with least squares fits on different
intervals of $u$, we found that for example by choosing $a_{id}=3/4$
and $b_{id}=1/5$, we obtain a reasonably good agreement between the
two curves over a large interval of $u$, as shown in Fig.
\ref{fig:Fu}. Even though details in the fitting procedure will
affect the obtained numbers to some extent, a common feature
observed in all reasonable fits is that the parameter $a_{id}$ is
always on the order of 1, while $b_{id}$ is almost an order of
magnitude smaller. Thus, this argument explains why the parameters
that were observed to be practicable in our previous studies
\citep{jaatinen2009} were so different from unity, suggested by the
naive Taylor expansion of the $\log$-term.

Further refinement to the parameters $a$ and $b$ could be obtained by studying the terms related to $\hat{C}(k)$ in Eq. \eqref{eq:largealphau}. However, as the expressions for $\rho$ and $n$ used in deriving Eqs. \eqref{eq:FuEOF} and \eqref{eq:largealphau}, we suggest using numerical fitting methods in order to achieve maximum accuracy. In order to find a functional, that gives such a field $n$ that best approximates the original DFT, the most obvious choice is to find such $a$ and $b$ that a numerical free energy minimization results in the same $u$ as obtained from the DFT, and the solid that exhibits this $u$ coexists with the liquid phase (given that is the case in the original DFT functional as well).

For dynamics of the EOF model, we use the form widely used in PFC studies, i.e.
\begin{equation}
\begin{split}
\label{eq:eofn} \frac{ \partial n}{\partial \tau} = \nabla^2 & \left( \frac{\delta F_{EOF}}{\delta n}\right) \\
= \nabla^2 &
\left[ \left(1-\hat{C}(k_m) \right)n  - \frac{a}{2}n^2 +
\frac{b}{3}n^3  + \right.
\\ & \left. E_S \left( \frac{k_m^2 + \nabla^2}{k_m^2}
\right)^2 n + E_B \left( \frac{k_m^2 + \nabla^2}{k_m^2}\right)^4n
\right],
\end{split}
\end{equation}
where $\tau$ is defined as earlier. Motivation to choosing this
equation of motion is that it is probably the minimum complexity
model satisfying the usual requirements for conserving the total
mass and evolving towards minimum of the free energy, that also
catches approximately the same dynamics as the DDFT in the
near-uniformity limit, for the relevant Fourier modes up to $k_m$
(linearized version of Eq. \eqref{eq:ddftn} is $\partial_\tau
\hat{n}(k) = -k^2 (1-\hat{C}(k))\hat{n}(k)$, while that of Eq.
\eqref{eq:eofn} would be the same, but with $\hat{C}$ replaced by
$\hat{C}_{EOF}$). For studying solidification, we believe the limit
of near-uniformity is the dominant factor affecting the
solidification front velocity, even though some details of the
dynamics on the solid side of the front may have a secondary effect
on the front propagation.

\subsection{Other phase field crystal models}

In addition to comparing the results of the EOF model with those of
the DDFT model, we will also compare their results to two other PFC
models presented in the literature. The first model we will include
is essentially the Swift-Hohenberg (SH) model used in almost all the
PFC studies up to date. Where the EOF model contains gradients up to
eighth order in the linear part of the free energy, the SH
formulation only contains gradients up to fourth order. The
procedure we use for obtaining parameters for the SH model such that
it could be used to model real parameters is essentially the one
introduced by Wu and Karma \citep{wu2007}. In the framework of the
present work, we may also view the SH-based approach an
approximation to the EOF, where $E_{B}=0$, and the parameters $a$
and $b$ are fitted through the same procedure as in the EOF. The
equation of motion for this model, which we will call the
fourth-order fit (FOF) for the remainder of this paper, becomes
\begin{equation}
\begin{split}
\label{eq:fofn} \frac{ \partial n}{\partial \tau} =  \nabla^2  &
\left[ \left(1-\hat{C}(k_m) \right)n  - \frac{a}{2}n^2 +
\frac{b}{3}n^3  + \right.
\\ & \left. E_S \left( \frac{k_m^2 + \nabla^2}{k_m^2}
\right)^2 n
\right],
\end{split}
\end{equation}
We note here that even though many formulations of the SH-based PFC
models do not include the third order term $-an^3/6$ in the free
energy, it has been shown that the model without that term can be
exactly recovered from Eq. \eqref{eq:fofn} after appropriate scaling
of the field variable and the parameters. \citep{jaatinen2010b}

In addition to the FOF and EOF models, another interesting PFC model
was proposed in the recent work of van Teeffelen \etal
\citep{teeffelen2009}. In the model they call PFC1, they start with
the DDFT, and approximate the function $\hat{C}(k)$ by expanding it
around $k_m$ in a fourth order power series, in a similar manner as
in the FOF model. However, as the excess part of the free energy in
this approach would not be sufficient to stabilize the solid at any
reasonable density \citep{jaatinen2009,teeffelen2009}, the excess
part of the free energy is multiplied by a scaling factor $\alpha$.
Thus, the equation of motion in this model becomes
\begin{equation}
\begin{split}
\label{eq:pfc1}
 \frac{ \partial n}{\partial \tau} & = \nabla^2 n - \\
& \alpha \nabla \cdot \left[ (1+n) \nabla \left\{ \left(
\hat{C}(k_m)- E_S \left( \frac{k_m^2 + \nabla^2}{k_m^2}\right)^2
\right) n \right\} \right].
\end{split}
\end{equation}
Using arguments based on a single mode approximation to the free
energy of the solid, van Teeffelen \etal come up with $\alpha=1.15$
for the case of colloids interacting via $r^{-3}$ potential which
they studied \citep{teeffelen2009}. In the present work, we will
utilize numerical fitting methods to find an $\alpha$, such that the
correct freezing point from the DFT is reproduced in the PFC1 model.
Van Teeffelen \etal also showed that this PFC1 model performs
slightly better than their "PFC2" model, which is based on FOF with
$a=b=1$ and $\alpha=1.15$, in reproducing the initial
crystallization velocities of the DDFT, arguing that the better
success is due to fewer approximations made in the derivation.
\citep{teeffelen2009} In the following, we shall see how the PFC1
model compares with EOF and FOF for the case under study in the
present paper.

\section{Static results and model parameters}
\label{sec:statics}

In order to find the freezing density of the fluid in the DFT model,
we have calculated the direct correlation function for the pair
potential defined by Eq. \eqref{eq:potential} at different densities
by using the well-known Ornstein-Zernike equation together with the
closure relation by Percus and Yevick (PY) \citep{percus1958}. At
each density, we then find the non-trivial minimum of the free
energy, in which the density field has a hexagonal structure (at
densities where it exists), by a similar free energy minimization
method as utilized in \citep{jaatinen2009}. Of the minimization
method, it is worthy of noting that unlike in most DFT studies
\citep{singh1991,teeffelen2009} (and like in \citep{jaatinen2009}),
we do not restrict the calculations to a perfect lattice (i.e. zero
vacancy concentration) constraint, for the reason that in a
dynamical simulation, it is not possible to control this issue
without modifying the free energy.

Repeating the procedure of finding the direct correlation function
and the solid minimum of free energy at many different reference
densities, we find that freezing occurs at $\tilde{\rho}_l=0.9156$,
because for that reference density, the minimum free energy of the
solid equals that of the liquid at $\rho=\rho_0$. That is also the
reference density $\rho_0$ that will be used for all the
calculations in the remainder of this paper. Compared with the
previously mentioned result from molecular dynamics simulations,
this result is an underestimation of the freezing density by
approximately 7 \%. This difference could be due to multiple sources
of error, including approximating $F_{xs}$ by Eq. \eqref{eq:Fxs},
the PY closure relation or not including the perfect lattice
constraint. However, for our purposes, the result is acceptable,
especially given that the width of the coexistence gap $\Delta
\rho^* = (\rho_s-\rho_l)/\rho_l \approx 2 \%$ \citep{broughton1982}
is reproduced well: Our DFT result is $\Delta \rho^* = 2.20 \%$.

\begin{figure}
\begin{center}

{\includegraphics[width=86mm]{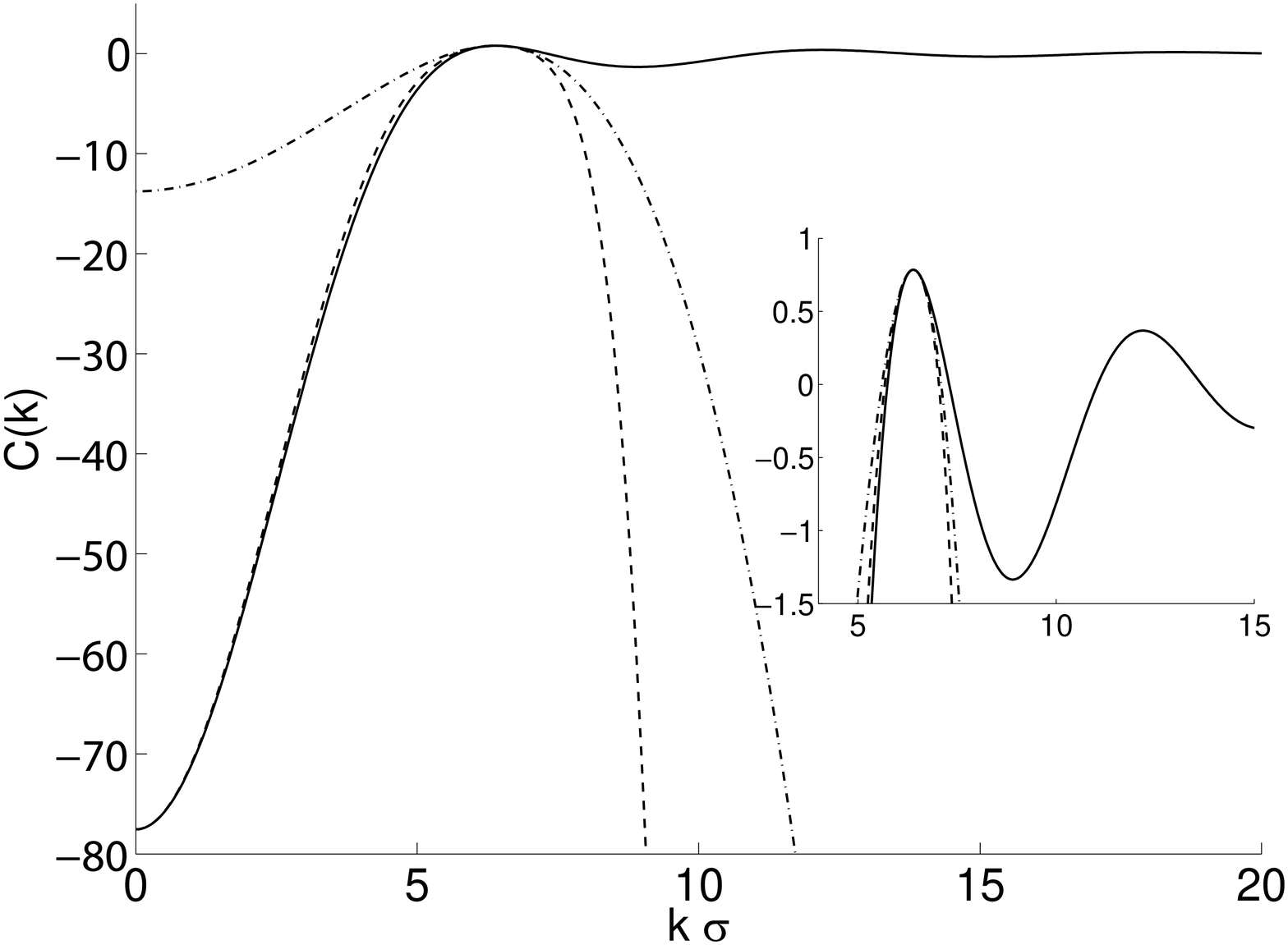}}

\caption{ Direct correlation function at $\tilde{\rho}_l=0.9156$ as
obtained from the PY closure relation (solid line), and the fitted
EOF (dashed line) and FOF (dashed-dotted line) expansions.
\label{fig:ck}}
\end{center}
\end{figure}

From the direct correlation function at freezing point, we find the
PFC model parameters $k_m=6.3965 /\sigma$, $\hat{C}(k_m)=0.7855$,
$E_S=14.5487$ and $E_B=63.7814$. The direct correlation function,
together with the expansions of EOF (Eq. \eqref{eq:cexpansion}) and
FOF (Eq. \eqref{eq:cexpansion} with $E_B=0$) are shown in Fig.
\ref{fig:ck}. As mentioned in the previous section, for the EOF and
FOF models, the parameters $a$ and $b$ are then defined such that
the solid phase coexists with the liquid phase at $\rho=\rho_0$, and
the amplitude of Fourier modes corresponding to the first star of
reciprocal lattice vectors of the solid phase
\begin{equation} \label{eq:us}
u_s = \int d{\bf r} n({\bf r}) e^{i {\bf G \cdot r}},
\end{equation}
where ${\bf G}$ is any reciprocal lattice vector from the first
star, equals that obtained from the DFT, $u_s=0.7914$. Based on
numerical iteration, these two constraints yield $a=0.8082$ and
$b=0.1388$ for the EOF, which are notably rather close to the values
of $a_{id}$ and $b_{id}$ presented in the previous section. For FOF,
the same fitting procedure results in only slightly different
numbers, $a=0.7812$ and $b=0.1438$. For PFC1, the parameter $\alpha$
is chosen to fulfill only first of the constraints for EOF and FOF,
i.e. that the solid coexists with the liquid at $\rho=\rho_0$. This
yields $\alpha=1.1934$.

As expected \citep{teeffelen2009,jaatinen2009}, the solid state
density profiles we find from the DFT are much more peaked around
the lattice sites. From all PFC models, we find the field $n$
resembles the one-mode approximation rather closely. The $n$ fields
found from EOF and FOF are very similar to each other, and to the
field obtained from the DFT through filtering the higher order
Fourier modes in the density. The solution from PFC1 differs from
the two other PFC models in that the amplitude of density
fluctuations is $u_s=0.2051$, which is smaller than in the DFT and
the other PFC models by about a factor four. The coexistence gap
$\Delta \rho^*$ is 1.57 \% in the EOF, 7.70 \% in the FOF and 0.68
\% in the PFC1. Comparing the coexistence gaps of the PFC models
with the previously mentioned results from molecular dynamics and
DFT, the EOF gives the closest, although not perfect result. In FOF,
too small a bulk modulus results in too large a coexistence gap
\citep{jaatinen2009}. In PFC1, the small $u_s$ and $\Delta \rho^*$
indicate that the transition from solid to liquid is a weaker first
order transition than in the two other models.

\begin{figure}
\begin{center}
{\includegraphics[width=86mm]{surfaces2D.eps}}
{\includegraphics[width=95mm]{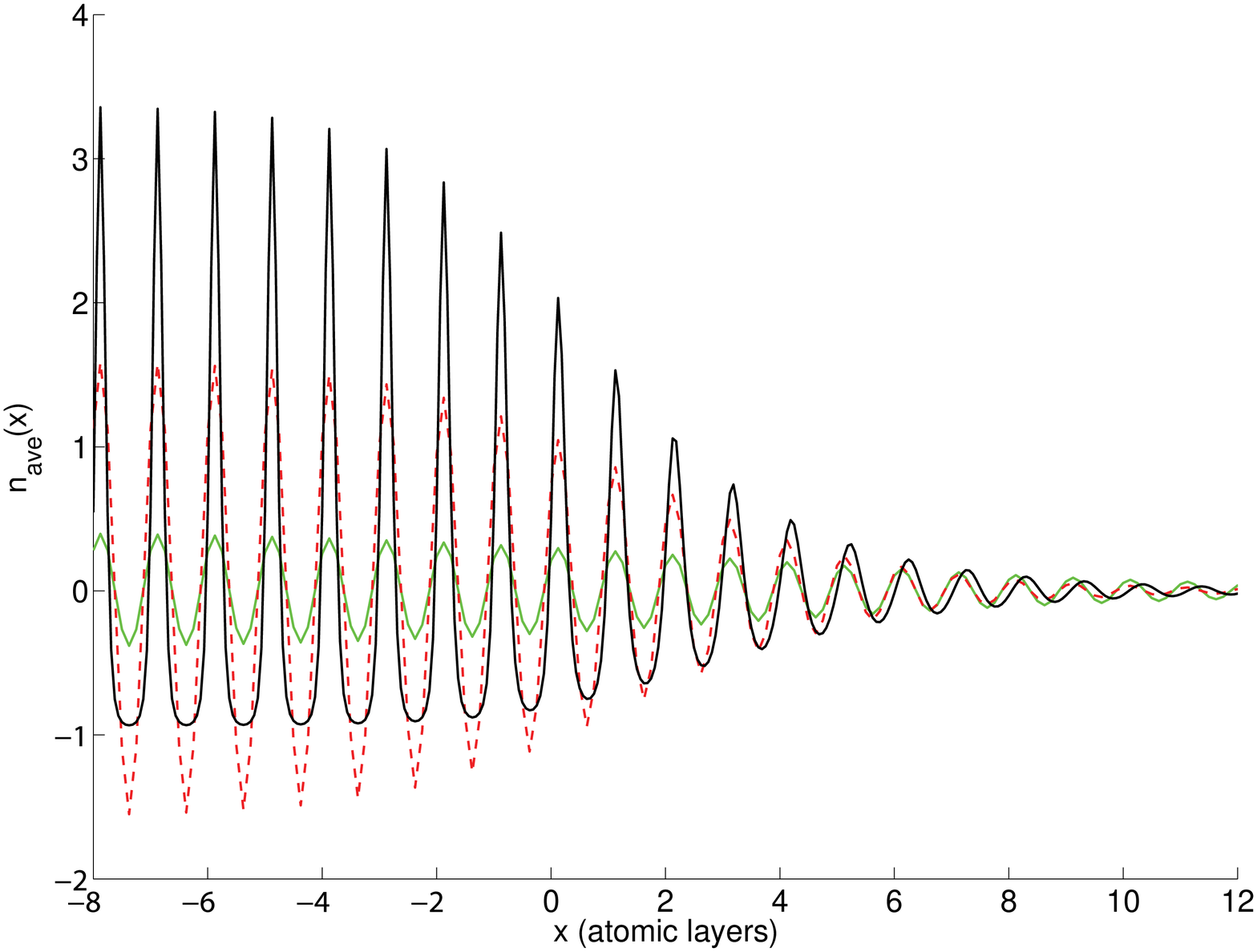}}

\caption{ (color online) On the top, we show density profiles of the
solid-liquid interface, from top to down, in the DFT, EOF, FOF and
PFC1 models. Darker shades of gray correspond to larger densities in
these images, with a scale such that the maximum of $n$ in each case
corresponds to black, and minimum to white. In the lowest image, we
show the field $n$ averaged in the direction parallel to the
interface, as a function of the spatial coordinate perpendicular to
the interface. Black line corresponds to DFT, red dashed line is EOF
and green line is PFC1 (FOF is not shown, because it would be
practically indistinguishable from EOF in this plot).
\label{fig:surf}}
\end{center}
\end{figure}

We have also studied properties of the solid-liquid interface in
these models in the close-packed $\left[ 10 \right]$ direction. Due
to scaling properties of the potential, the interfacial free energy
of the system is given by
\begin{equation}
\label{eq:gamma} \Gamma = \frac{k_BT}{\sigma} \left(
\frac{k_BT}{\epsilon}\right)^{1/6} \tilde{\Gamma},
\end{equation}
where $\tilde{\Gamma}$ is the dimensionless interfacial free energy
in rescaled units. Initializing a system with a slab of solid at
$\tilde{\rho}_s$ in the middle of a liquid at $\tilde{\rho}_l$, we
find after numerical free energy minimization $\tilde{\Gamma} =
0.234$ in the DFT. From the EOF and FOF we find almost exactly the
same interfacial energies, i.e. $\tilde{\Gamma} = 0.222$ in the EOF
and $\tilde{\Gamma} = 0.223$ in the FOF. These results also agree
rather well with the DFT result. On the other hand, from PFC1 we
find $\tilde{\Gamma} = 0.0086$, which is more than an order of
magnitude smaller than in all the other models. Density profiles of
the interface layer in the different models are shown in Fig.
\ref{fig:surf}. It can be seen that in PFC1, the interface layer is
considerably wider than in all the other models, while the interface
widths in EOF and FOF are very similar to that in the DFT. We are
not aware of any computer simulation predictions for the surface
free energy of $r^{-12}$ disks.

As an aside, we note that the justification for $a$ and $b$ being
different from unity presented in the previous section is not the
only one published. Berry \etal \citep{berry2008} have noted that a
local $n^3$ term can also be justified by considering the $k=k'=0$
contribution from a third order term in the density expansion of
$F_{xs}$. Technically, this is equivalent to assuming that the
density field is slowly varying compared with the range of three
body correlations. If such term is included in Eq. \eqref{eq:Fxs},
and the logarithm is expanded in a Taylor series, it is
straightforward to derive an explicit expression for $a$,
\begin{equation}
\label{eq:a3rd} a = 1 + \rho_0^2\hat{c}^{(3)}(0,0),
\end{equation}
where $\hat{c}^{(3)}(0,0)$ is the $k=k'=0$ mode of the three body
direct correlation function. Using the Ornstein-Zernike relation
with Percus-Yevick closure, $\hat{c}^{(3)}(0,0)$ can be calculated
by noting that it is related to the $k=0$ mode of the two body
direct correlation function through the sum rule
\begin{equation}
\label{eq:c3} \hat{c}^{(3)}(0,0) = \frac{\partial
\hat{c}^{(2)}(0;\rho_0)}{\partial \rho_0}.
\end{equation}
The prediction for $a$ we obtain from this approach is $a \approx
-410$, which is not only large in terms of absolute value, but also
has the wrong sign with respect to stabilizing the solid phase.
Similar consideration for the fourth-order term $b$, considering
$k=k'=k''=0$ contribution from the four-body correlation term,
yields $b \approx 1.35 \times 10^3$. Based on these considerations,
we conclude that our values for $a$ and $b$ cannot be justified in
terms of higher order correlations.

\section{Dynamical simulations}
\label{sec:dynamics}

The solidification front dynamics in the DDFT and the different PFC
models were studied by growing a hexagonal crystal from an
undercooled liquid (i.e. a liquid with an initial density
$\rho_i>\rho_l$) in the $[10]$ direction. In the direction
perpendicular to the solidification front propagation, the size of
the array in our computations is exactly one interparticle spacing,
and periodic boundary conditions are used. Due to the periodic
boundaries, our simulations represent an infinitely wide crystal
seed that propagates into the liquid. That the size of our
simulation box is only one interparticle spacing in the $[01]$
direction means that no instabilities that could roughen the surface
are allowed. Initial condition for the DDFT, EOF and FOF simulations
is such that eight monolayers of perfect solid are placed in the
middle of the undercooled liquid, with a slight smoothing in the
boundary of solid and liquid phases, while for PFC1, we had to use a
crystal seed of 12 monolayers in order to initialize the growth
process at even the smallest undercoolings studied. Once the
simulation starts, the solid seed grows in both directions, and we
measure its position as a function of time. Position of the surface
is defined as the point where a line drawn through the local maxima
of the density corresponding to the solid particles reaches one half
of its maximum value. In the direction of growth, the length of our
array was usually 512 interparticle spacings.

For numerically integrating the EOF and FOF models, we use the
well-known semi-implicit operator splitting method
\citep{semiimplicit} with Fast Fourier transforms. For spatial
resolution, we use $\Delta x=\sqrt{3}d_{nn}/16$ and $\Delta
y=d_{nn}/16$ ($d_{nn}$ is the nearest-neighbor distance), time step
is $\Delta \tau=10^{-3}$, and the Laplace operator is discretized in
$k$-space as $\nabla^2_{\bf k}=-k^2$. For the DDFT, we employ a
similar procedure, by treating the $\nabla^2 n$ term in Eq.
\eqref{eq:ddftn} implicitly and the term related to $F_{xs}$
explicitly. The non-local term is evaluated in $k$-space and
derivatives in $x$ and $y$ directions are discretized in $k$-space
as $ik_{x,y}$. For DDFT, in order to resolve the sharp density
peaks, the linear spatial resolution of the PFC is doubled, i.e.
$\Delta x=\sqrt{3}d_{nn}/32$ and $\Delta y=d_{nn}/32$. The time step
we use for DDFT is $\Delta \tau = 10^{-3}$ in the regime of low
undercooling. In the regime of high undercooling we found that
retaining the numerical accuracy required us to decrease the time
step to $\Delta \tau = 10^{-4}$, which is smaller than we utilized
in EOF and FOF models by an order of magnitude. This, together with
the difference in spatial resolution, means that using our methods,
simulations with the DDFT are approximately two orders of magnitude
slower than with the EOF and FOF models. For PFC1, we modified the
method used for DDFT such that the implicitly treated part is
$\nabla^2(1-\alpha C_{PFC1}(\nabla^2))n$, leaving $\alpha \nabla
\cdot n \nabla( C_{PFC1}(\nabla^2)n)$ treated explicitly. Even
though for PFC1 this modification brought great advantage in
numerical stability, handling the non-linear part still involves
explicit evaluation of sixth derivative. Therefore, we found the
PFC1 to be numerically most unstable among the models studied. In
order to ease the requirement this model places on the time step, we
dropped the spatial resolution perpendicular to growth below that
used in other PFC models, to $\Delta y=d_{nn}/8$, which we did not
find to have any profound effect on any results predicted by the
model. However, this only allowed us to utilize time steps that are
one order of magnitude smaller than in the DDFT, making the PFC1
numerically approximately equally demanding to DDFT using current
methods. In addition to the differences in $\Delta x$, $\Delta y$
and $\Delta \tau$ between the models, we note that progressing a
single time step in DDFT and PFC1 models requires a total of five
Fourier (or inverse Fourier) transforms, where in EOF and FOF, only
three are required.

If the undercooling is small, such that $\rho_i<\rho_s$, then
formation velocity of the solid (whose density is always at least
$\rho_s$ to be stable) is expected to be limited by the diffusion of
mass to the interface, in an analogous manner to the more commonly
considered case where growth of the solid is limited by transport of
heat away from the surface \citep{lowen1992}. Density of the solid
seed in these simulations is chosen to be that of the solid
coexisting with the liquid. As the planar solidification front
propagates, the layer through which diffusion must take place
widens, and therefore one expects the solidification front to
propagate as $x \sim \tau^{0.5}$. Diffusion controlled growth in the
PFC model has been previously studied by Tegze \etal
\citep{tegze2009}, but to our knowledge, no studies of the subject
utilizing a non-local DDFT have been published. On the other hand,
when density of the liquid from which the solid is formed exceeds
$\rho_s$, propagation of the solidification front does not require
diffusion of additional mass to the surface. Therefore, one expects
the front to propagate with a constant velocity,  i.e. $x \sim t$,
that depends on the attachment rate of particles on the interface.

\begin{figure}
\begin{center}
{\includegraphics[width=86mm]{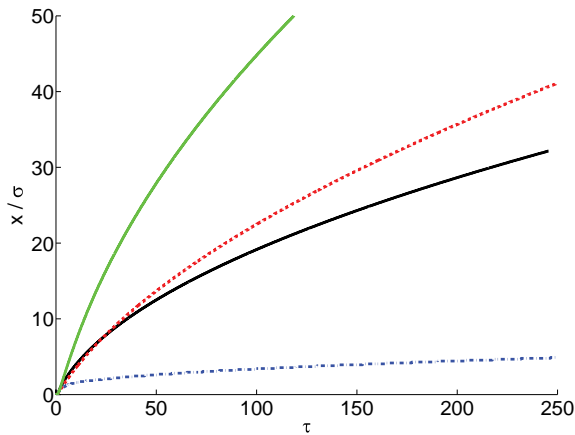}} \caption{(color online)
Interface position as a function of time when $\rho_i=1.0044\rho_l$.
Black solid line is the result obtained from the DDFT, red dashed
line is from the EOF, blue dash-dotted line is from the FOF and
green solid line is from the PFC1. \label{fig:xtdiff}}
\end{center}
\end{figure}

\begin{figure}
\begin{center}
{\includegraphics[width=86mm]{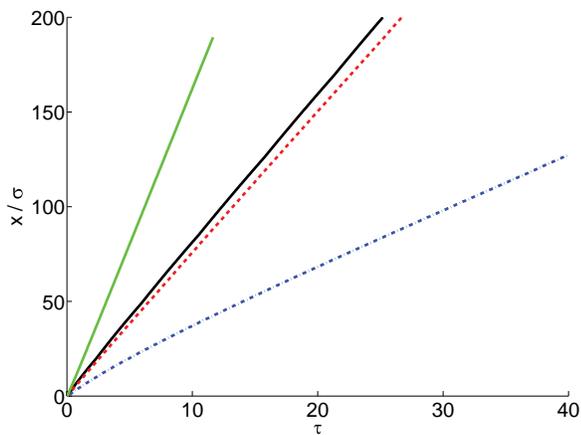}} \caption{Interface
position as a function of time when $\rho_i=1.088\rho_l$. Different
lines are as in Fig. \ref{fig:xtdiff}. \label{fig:xtkin}}
\end{center}
\end{figure}

The different regimes of growth, as well as differences between the
different models, are illustrated in Figs. \ref{fig:xtdiff} and
\ref{fig:xtkin}. In Fig. \ref{fig:xtkin} we show the interface
position as a function of time from all models, for a case where
$\rho_i=1.0044\rho_l$, which is in the $\Delta<1$ regime of all
models. At the very beginning of the solidification process,
kinetics define how fast the particles attach to the interface from
the liquid with a density $\rho_i$. During the initial stages, a
depletion layer is formed in front of the liquid. As time goes on,
width of the depletion layer increases, and $x(\tau)$ approaches the
expected $x \sim \tau^{0.5}$ behavior. By close inspection of Fig.
\ref{fig:xtdiff} it can be seen that at the very beginning of the
process, interface motion in the PFC1 model is slightly faster than
in the DDFT. On the other hand in the EOF and FOF models, the
initial interface motion is slightly slower than in DDFT, indicating
more restriction to growth due to interface kinetics. In the EOF and
FOF models, the initial velocities are strikingly similar. On the
other hand, Fig. \ref{fig:xtkin} shows interface positions as a
function of time for an initial density of $\rho_i=1.088\rho_l$,
which is in the $\Delta>1$ regime of all models. Again, in the
initial stages of solidification, a depletion layer is formed in
front of the moving interface. However in this case the width of the
depletion layer, and thus the propagation velocity of the interface,
quickly approach constant values, and therefore the growth seems
linear.

\begin{figure}
\begin{center}
{\includegraphics[width=86mm]{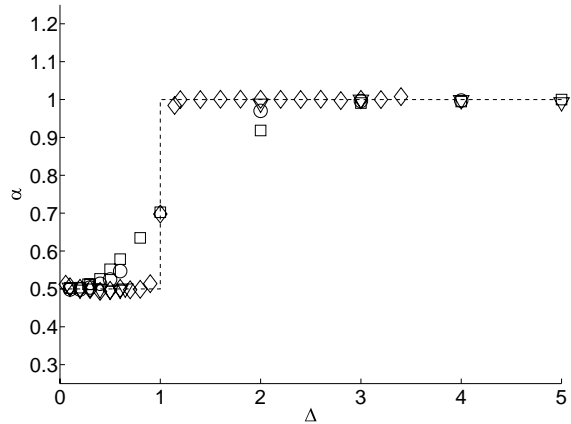}}
\caption{\label{fig:exponents} Growth exponents obtained from the
different models.  Circles are results from DDFT model, squares from
EOF, diamonds from FOF and triangles from PFC1, while dashed line
shows the ideal behavior.}
\end{center}
\end{figure}

\begin{figure}
\begin{center}
{\includegraphics[width=86mm]{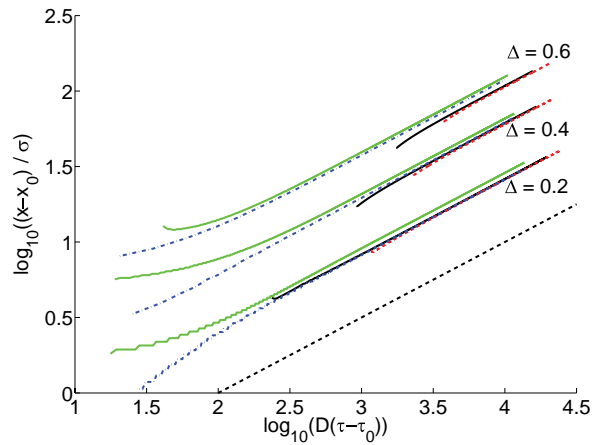}} \caption{(color online)
Logarithmic plot of rescaled interface positions as a function of
rescaled time from different models in $\Delta \leq 0.6$ regime. The
black dashed line shows the expected slope, while other lines are
labeled as in Fig. \ref{fig:xtdiff}. \label{fig:xtdiff2}}
\end{center}
\end{figure}

In order to quantify the detection of the different growth regimes,
we have fitted a power law growth function,
\begin{equation}
\label{eq:plaw} x = x_0 + c (\tau-\tau_0)^\alpha,
\end{equation}
in the surface positions as a function time resulting from the
different models at different initial densities. The first quarter
of the $x(\tau)$ data is ignored in these fits in order to minimize
the effect from the initial stages, while still obtaining a fairly
robust fit for the four free parameters. The exponents $\alpha$
resulting from these fits are shown in Fig. \ref{fig:exponents}. The
growth exponents obtained from the FOF and PFC1 models are in
general close to the ideal results (even at unit undercooling, the
observed exponent $\alpha \approx 0.70$ from FOF is close to the
expected anomalous exponent $\alpha=2/3$
\citep{lowen1992,marder1992}), whereas in DDFT and EOF models, the
transition from the diffusion controlled to kinetics controlled
regime seems more continuous. The discrepancy between observed and
expected exponents is due to insufficient simulation time for the
formation of the steady-state depletion layer in DDFT and EOF
models. The discrepancy is most evident in the EOF model. This is
most likely because the combination of small coexistence gap, fast
diffusion in the solid, and relatively slow interfacial kinetics
makes the formation of a quasi-steady-state depletion layer in the
EOF slowest among the models.

\begin{figure}
\begin{center}
{\includegraphics[width=86mm]{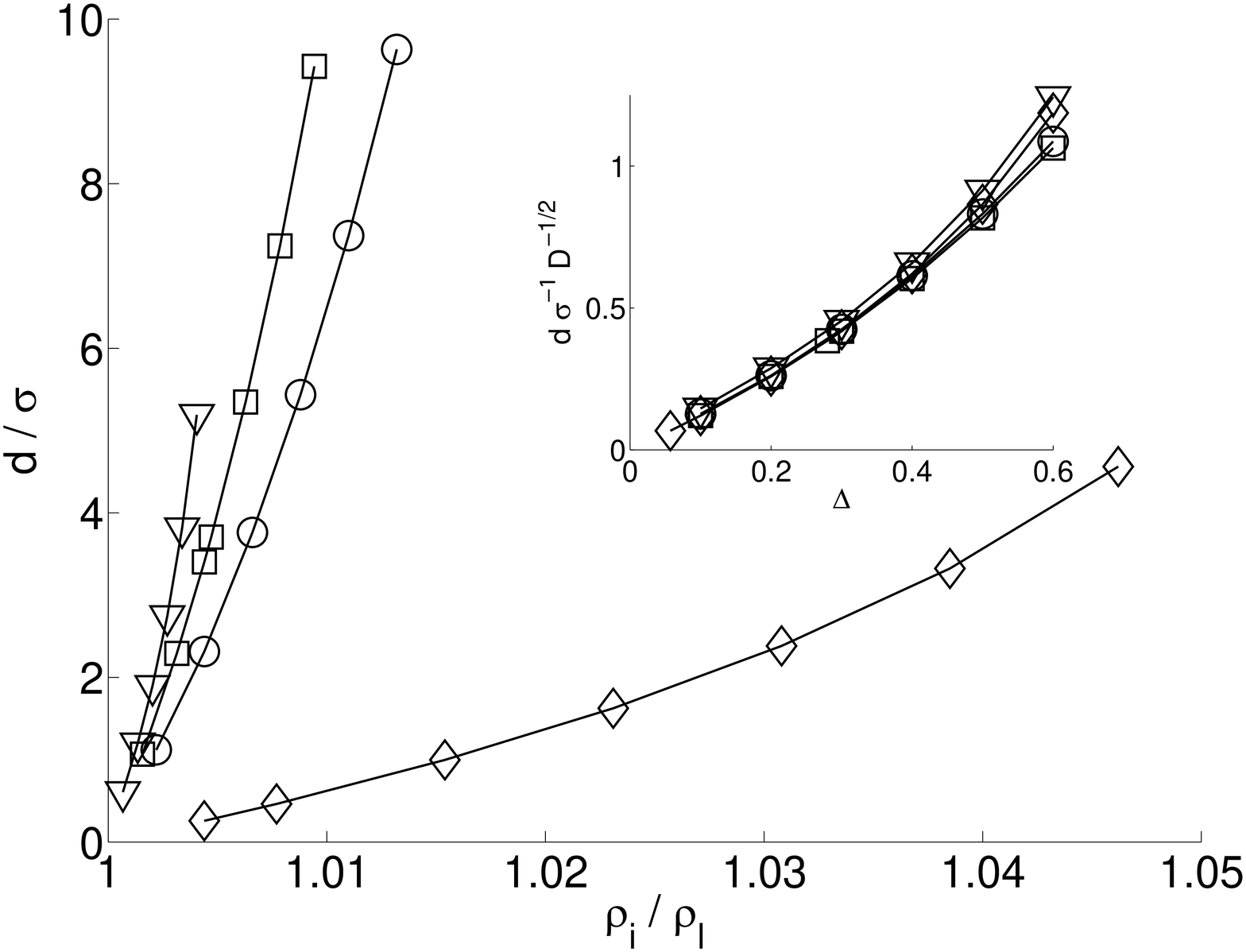}} \caption{Growth rates
$d$ obtained in the diffusion controlled regime. Circles are results
from DDFT model, squares from EOF, diamonds from FOF and triangles
from PFC1, with lines connecting the symbols. Inset shows scaled
data. \label{fig:diffgrowth}}
\end{center}
\end{figure}

In the $\Delta<1$ regime, we have quantified the effect of the
initial density on the quasi-steady-state velocity of front
propagation by fitting the $x(\tau)$'s resulting from different
models for different $\rho_i$'s with the expected growth law,
\begin{equation}
\label{eq:diffgrowth} x = x_0 + d (\tau - \tau_0)^{\frac{1}{2}},
\end{equation}
ignoring the early stages where $x\sigma^-1<50$. Eq.
\eqref{eq:diffgrowth} is mathematically equivalent to Eq.
\eqref{eq:plaw} with $\alpha$ set to 0.5. Despite the previously
mentioned discrepancies between the observed and now pre-set growth
exponents, we are able to obtain good fit with Eq.
\eqref{eq:diffgrowth} for all models in the regime $\Delta \leq
0.6$, as shown in Fig. \ref{fig:xtdiff2}. The $d$ resulting from
these fits are shown in Fig. \ref{fig:diffgrowth}. Beyond $\Delta =
0.6$, we have less confidence in having reached close enough to a
steady-state diffusion-controlled growth regime, especially in the
EOF model, and therefore the data is only shown for undercoolings up
to $\Delta = 0.6$.

It is seen in Fig. \ref{fig:diffgrowth} that among the PFC models,
the EOF in general gives the closest agreement with the DDFT. Due to
scaling properties of the problem, if the growth were purely
diffusion limited, one would expect the position of the interface to
depend on the dimensionless undercooling $\Delta$ and $\sqrt{D
\tau}$, where $D=1-\hat{C}(0)$ (for PFC1, multiply $\hat{C}(0)$ by
$\alpha$) is the effective diffusion constant in a given model (in
the limit of small, long-wavelength density fluctuations, with this
definition, all the models studied reduce to the diffusion equation
$\partial_\tau n = D \nabla^2 n$). Such a scaling law for the growth
rates is illustrated in the inset of Fig. \ref{fig:diffgrowth},
where we show that the $d$'s scaled by $\sqrt{D}$ as a function of
$\Delta$ in DDFT, EOF and FOF models follow the same curve,
indicating that differences in microscopic details of those models
are unimportant in determining the front velocity in the diffusion
controlled regime. Results from PFC1 lie slightly above those from
the other models in the rescaled plot, which we believe is most
likely a result of numerical inaccuracy. These scaling properties of
the problem are the reason why the EOF, which has exactly the same
$D$ and close to the same $\Delta \rho^*$ as DDFT, reproduces the
result of the DDFT with higher accuracy than the FOF, which results
in a $d$ that is approximately an eighth of the result obtained from
the DDFT for the same $\rho_i$. The scaling argument also suggests
that the close agreement of PFC1 to EOF and DDFT in the small
density limit probably results from a cancelation of errors due to
the smaller $D$ and smaller $\Delta \rho^*$ in the PFC1 model.

\begin{figure}
\begin{center}
{\includegraphics[width=86mm]{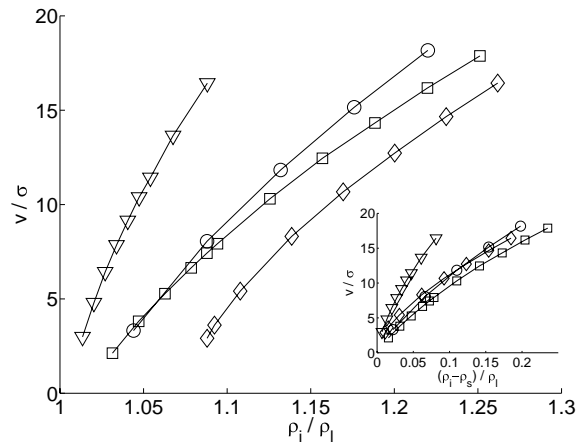}} \caption{Growth
velocities $v$ obtained in the kinetics controlled regime. Symbols
as in Fig. \ref{fig:diffgrowth}. \label{fig:kingrowth}}
\end{center}
\end{figure}

In the regime where $\Delta > 1$, the results have been fitted with
a linear growth law,
\begin{equation}
\label{eq:diffusiongrowth} x = x_0 + v \tau,
\end{equation}
where the effect of the initial transient was removed by ignoring
data for which $x<100 \sigma$. The resulting $v$ as a function of
$\rho_i$ from all the models are shown in Fig. \ref{fig:kingrowth}.
As expected, the front velocity increases as the initial density is
increased in all the models. It is also apparent that for any given
initial density, the velocity obtained from the EOF is the closest
approximation to the DDFT among the present PFC models. If the
density axis is rescaled by subtracting the density of the solid
coexisting with the liquid, the results from FOF seem to agree with
DDFT practically as well as those from the EOF, as shown in the
inset of Fig. \ref{fig:kingrowth}. On the other hand, the velocities
observed in PFC1 model seem to be a significant overestimation when
compared with the results from all the other models studied, even
after rescaling the densities. This suggests that the smaller
amplitude of density fluctuations and wider crystal-melt surface
result in a kinetic barrier which is somewhat smaller than in the
other models.

\section{Conclusions}
\label{sec:conclusion}

We have presented a new way to derive the eighth-order phase field
crystal model (EOF) from the density functional theory of classical
systems. The model was applied to study solidification front
dynamics in a two-dimensional ensemble of particles interacting via
$r^{-12}$ potential. Predictions from the EOF were compared with
similar predictions from dynamical density functional theory (DDFT)
of Marconi and Tarazona, and two previously presented phase-field
crystal (PFC) models. For the static properties of the system in
these models, we find that the DFT predicts freezing of the
$r^{-12}$ disks at a density that is about $7$ \% lower than seen in
molecular dynamics simulations. From the PFC models studied, we find
that the EOF gives the most accurate description of the static
properties of the material under study. By studying crystal growth
in the diffusion controlled regime, we find that the EOF gives the
best agreement with DDFT among the phase field crystal models, due
to the most accurate description of liquid diffusion constant and
solid-liquid coexistence gap in the model. In the regime of
interface kinetics controlled growth, we again find the EOF gives
closest agreement to the DDFT for all initial densities, although if
the initial density is rescaled by the melting point of the solid,
the fourth order fitting scheme slightly outperforms the EOF. These
results suggest that among the PFC models studied, the EOF gives the
closest approximation to the DDFT. This implies that the EOF is a
good candidate for a model to be used for atomistic scale
simulations of the growth of two dimensional hexagonal crystals of
Brownian particles, at least in the absence of an external field. In
the presence of an external field, a further study would be required
to quantify the response in the different models. It should also be
noted that while the current study has considered a simple
two-dimensional problem, a similar study of the growth of a three
dimensional crystal could also quantify the differences in
anisotropy of the different phenomena in the models.

\section{Acknowledgements}
We wish to thank prof. K. R. Elder for insightful discussions
concerning this work. This work has been supported in part by the
Academy of Finland through its Center of Excellence COMP grant and
Tekes through its MASIT33 project. A.J. acknowledges support from
the Finnish Foundation for Technology Promotion. We also wish to
thank CSC-Scientific Computing Ltd. for computational resources.





\begin{thebibliography}{100}

\bibitem{elder2002} K. R. Elder, M. Katakowski, M. Haataja and M. Grant,
Phys. Rev. Lett. {\bf 88}, 245701 (2002)

\bibitem{elder2004} K. R. Elder and M. Grant, Phys. Rev. E {\bf 70}, 051605 (2004)

\bibitem{emmerich2008} H. Emmerich, Adv. Phys. {\bf 57}, 1 (2008)

\bibitem{stefanovic2006} P. Stefanovic, M. Haataja and N. Provatas, Phys. Rev. Lett. {\bf 96}, 225504 (2006)

\bibitem{berry2006} J. Berry, M. Grant and K. R. Elder, Phys. Rev. E 73, 031609 (2006)

\bibitem{tegze2009} G. Tegze, L. Granasy, G. I. Toth, F. Podmaniczky, A. Jaatinen, T. Ala-Nissila and T.
Pusztai, Phys. Rev. Lett. {\bf 103}, 035702 (2009)

\bibitem{teeffelen2009} S. van Teeffelen, R. Backofen, A. Voigt and
H. L\"owen, Phys. Rev. E {\bf 79}, 051404 (2009)

\bibitem{achim2006} C. V. Achim, M. Karttunen, K. R. Elder, E. Granato, T. Ala-Nissila and S. C.
Ying, Phys. Rev. E {\bf 74}, 021104 (2006)

\bibitem{achim2008} C. V. Achim, M. Karttunen, K. R. Elder, E. Granato, T. Ala-Nissila and S. C.
Ying, J. Phys. Conference Series {\bf 100}, 072001 (2008)

\bibitem{achim2009} C. V. Achim, J. A. P. Ramos, M. Karttunen, K. R. Elder, E. Granato, T. Ala-Nissila, and S. C.
Ying, Phys. Rev. E {\bf 79}, 011606 (2009)

\bibitem{hansen2006} J.-P. Hansen and I. R. McDonald, \emph{Theory of Simple
Liquids}, 3$^{\mathrm{rd}}$ Edition (Academic, Amsterdam, 2006)

\bibitem{marconi1999}
U. M. B. Marconi and P. Tarazona, J. Chem. Phys. {\bf 110}, 8032
(1999)

\bibitem{ramakrishnan1979} T. V. Ramakrishnan and M. Yussouff,
Phys. Rev. B {\bf 19}, 2775 (1979)

\bibitem{singh1991}
Y. Singh, Phys. Rep. {\bf 207}, 351 (1991)

\bibitem{elder2007} K. R. Elder, N. Provatas, J. Berry, P. Stefanovic and M.
Grant, Phys. Rev. B {\bf 75}, 064107 (2007)

\bibitem{wu2007} K-A. Wu and A. Karma,
Phys. Rev. B {\bf 76}, 184107 (2007)


\bibitem{jaatinen2009} A. Jaatinen, C. V. Achim, K. R. Elder and T.
Ala-Nissila, Phys. Rev. E {\bf 80}, 031602 (2009)

\bibitem{jaatinen2010} A. Jaatinen, C. V. Achim, K. R. Elder and T.
Ala-Nissila, Technische Mechanik {\bf 30}, 169 (2010)

\bibitem{toth2010}G. I. T\'oth, G. Tegze, T. Pusztai, G. T\'oth and L
Gr\'an\'asy, J. Phys.: Condens. Matter {\bf 22} 364101 (2010)


\bibitem{lowen1996}H. Lowen, Phys. Rev. E {\bf 53}, R29 (1996)

\bibitem{broughton1982}J. Q. Broughton, G. H. Gilmer and J. D. Weeks, Phys. Rev. B {\bf
25}, 4651 (1982)

\bibitem{archer2004} A. J. Archer and R. Evans, J. Chem. Phys. {\bf 121}, 4246 (2004)

\bibitem{ramos2010} J. A. P. Ramos, E. Granato, S. C. Ying, C. V. Achim, K. R. Elder, and T. Ala-Nissila
Phys. Rev. E {\bf 81}, 011121 (2010)


\bibitem{teeffelen2007} S. van Teeffelen, C. N. Likos, and Hartmut
Lowen, Phys. Rev. Lett. {\bf 100}, 108302 (2008)

\bibitem{jaatinen2010b} A. Jaatinen and T. Ala-Nissila, J. Phys.: Condens. Matter {\bf 22}, 205402 (2010)

\bibitem{percus1958} J. K. Percus and G. J. Yevick, Phys. Rev. {\bf 110}, 1–13 (1958)

\bibitem{berry2008}J. Berry, K. R. Elder and M. Grant, Phys. Rev. E {\bf 77}, 061506 (2008)

\bibitem{semiimplicit}G. Tegze, G. Bansel, G. I. T\'oth, T. Pusztai, Z.
Fan and L. Granasy, J. Comp. Phys. {\bf 228}, (2009); B. P.
Vollmayr-Lee and Andrew D. Rutenberg, Phys. Rev. E {\bf 68}, 066703
(2003); J. Zhu, L-Q. Chen, J. Shen and V. Tikare, Phys. Rev. E {\bf
60}, 3564 - 3572 (1999)

\bibitem{lowen1992} H. L\"owen, J. Bechhoefer and L. S. Tuckerman,
Phys. Rev. A {\bf 45}, 2399 (1992)

\bibitem{marder1992} M. Marder, Phys. Rev. A {\bf 45}, R2158 (1992)











\end{thebibliography}
\end{document}